\documentclass[pre,twocolumn,superscriptaddress]{revtex4}

\def\be{\begin{equation}}
\def\ene{\end{equation}}

\begin{document}

\title{Physical Pictures of Transport in Heterogeneous Media:
Advection-Dispersion, Random Walk and Fractional Derivative Formulations}

\author{Brian Berkowitz}
\email{brian.berkowitz@weizmann.ac.il}
\affiliation{Department of Environmental Sciences and Energy Research,
Weizmann Institute of Science, 76100 Rehovot, ISRAEL}
\author{Joseph Klafter}
\email{klafter@post.tau.ac.il}
\affiliation{School of Chemistry, Tel Aviv University,
69978 Tel Aviv, ISRAEL}
\author{Ralf Metzler}
\email{metz@nordita.dk}
\affiliation{Department of Physics, Massachusetts Institute of
Technology, 77 Massachusetts Ave. Rm 12-109, Cambridge, MA 02139}
\affiliation{NORDITA, Blegdamsvej 17, DK-2100 Copenhagen {\O}, DENMARK}
\author{Harvey Scher}
\email{harvey.scher@weizmann.ac.il}
\affiliation{Department of Environmental Sciences and Energy Research,
Weizmann Institute of Science, 76100 Rehovot, ISRAEL}


\begin{abstract}
The basic conceptual picture and theoretical
basis for development of transport equations in porous media are examined.
The general form of the governing equations is derived for
conservative chemical transport in heterogeneous geological
formations, for single realizations and for ensemble averages
of the domain. The application of these transport
equations is focused on accounting for the appearance of non-Fickian
(anomalous) transport
behavior. The general ensemble-averaged transport equation is shown to be
equivalent
to a continuous time random walk (CTRW) and reduces to the conventional
forms of the
advection-dispersion equation (ADE) under highly restrictive conditions.
Fractional derivative formulations of the transport equations, both 
temporal and
spatial, emerge as special cases of the CTRW. In particular, the use 
in this context of
L{\'e}vy flights is critically examined. In order to determine 
chemical transport in
field-scale situations, the CTRW approach is generalized to 
non-stationary systems. We
outline a practical numerical scheme, similar to those used with 
extended geological
models, to account for the often important effects of unresolved 
heterogeneities.
\end{abstract}

\maketitle

\section*{1. Introduction}

Quantification of chemical transport mediated by flow fields
in strongly heterogeneous geological environments has
received an inordinate amount of attention over the last
three decades, and a vast literature dealing with the subject has developed
(see, e.g., the recent reviews in {\it Dagan and Neuman} [1997]).
Existing modeling approaches are generally based on various
deterministic and stochastic forms of the advection-dispersion
equation (ADE); the former include conditioning the domain of
interest by known heterogeneity structures, while the latter
include Monte Carlo, perturbation and spectral analyses.
A major feature of transport, particularly in more heterogeneous
domains, is the appearance of
``scale-dependent dispersion'' [e.g., {\it Gelhar et al.}, 1992].
Contrary to the fundamental
assumptions underlying use of the classical ADE (which
assumes a constant flow field and dispersion coefficients),
the very nature of
the dispersive transport seems to change as a function of
time or distance traveled by the contaminant.
Such scale-dependent behavior,
also sometimes referred to as ``pre-asymptotic'', ``anomalous'' or
``non-Gaussian'', is what we shall refer to as ``non-Fickian" transport.

Efforts to quantify non-Fickian transport have focused on more general
stochastic ADE's with, e.g., spatially varying velocity
fields. Stochastic analyses have provided substantial
insight into the dispersion process. They have been shown,
through application to well-documented field experiments,
to provide predictions of the temporal variation
of the first and second order moments of tracer plumes
in geological formations characterized by
relatively small degrees of heterogeneity (e.g., the Cape
Cod site [{\it Garabedian et al.}, 1991]).
Other variations based on the
classical ADE have also received attention; these include
``patch'' solutions which include an empirical
time- or space-dependent dispersivity, and mobile-immobile
and multirate diffusion type models [e.g., {\it Haggerty and
Gorelick}, 1995; {\it Harvey and Gorelick}, 2000].
However, the vast majority of these models assume, either explicitly
or implicitly, an underlying Fickian transport behavior at some
scale [e.g., {\it Sposito et al.}, 1986; {\it Rubin}, 1997].
Also, many of these approaches are based on perturbation theory,
and they are therefore limited
to porous media in which the variance of the log hydraulic
conductivity is small.

Other non-local formulations that do not invoke a Fickian
transport assumption have been hypothesized
and/or developed from various mathematical formalisms
[e.g., {\it Zhang}, 1992; {\it Glimm et al.}, 1993;
{\it Neuman}, 1993; {\it Deng et al.}, 1993;
{\it Cushman et al.}, 1994; {\it Dagan}, 1997].
These formalisms, in general, are founded on a fundamental separation
between advective and dispersive mechanisms; they yield
solutions (for the concentration) that result in definition of a
dispersion tensor that is usually formulated in
Fourier-Laplace space, whose inversion is difficult
to treat and/or apply.

Practical application of these models, to quantify the
full evolution of a migrating contaminant plume, has
not yet been achieved. In fact, the overwhelming emphasis of
these various studies has been limited to moment characterizations of
tracer plume migration, and/or to determination of the
``macrodispersion'' parameter. The complete solutions are not
analytically tractable, and their practical utility
remains largely undemonstrated.

The difficulty in capturing the complexities
of tracer plume migration patterns suggests that local, small-scale
heterogeneities cannot be neglected.
Evidently, these unresolvable heterogeneities contribute
significantly to the occurrence of non-Fickian transport.
The apparent existence  of hydraulic conductivity fields with coherence
lengths that vary over many scales suggests that temporal, as well as
spatial issues must be considered in any mathematical formulation.
Coupled to this problem is the lack of clarity of how best to use
field observations to reduce the inevitable uncertainties of
the model. Frequently, the latter issue involves
the interplay between ensemble averaging (probabilistic approaches)
and spatial scales of resolution of non-stationary geological
features.

In this paper, we re-evaluate the basic conceptual picture of tracer
migration in heterogeneous media. We derive the general form
of the governing equations for conservative chemical transport in
heterogeneous geological formations, for single realizations
and for ensemble averages of the domain.
We emphasize quantification of non-Fickian transport
behavior, and show that a
general form of the ensemble-averaged transport equation
is a continuous time random walk (CTRW).
In this framework, we show that non-Fickian transport
results from the inapplicability of the central limit
theorem to capture the distribution of particle transitions
(detailed in the next section).
Fractional derivative formulations of the transport equations,
both temporal and spatial, are seen to emerge from another
set of conditions, and are therefore special cases of the CTRW.
We then focus on quantifying transport in non-stationary media,
and discuss how best to deal with the coupled problem of
integrating ensemble averaging with information on non-stationarity
at various scales of resolution.

\section*{2. Governing Transport Equations for Heterogeneous Media}

\subsection*{2.1. Physical Framework of the Transport Equations}

Contaminants disperse as they migrate within the flow field
of the geological maze we call an aquifer. At the outset one must choose an
underlying physical model of this process.
Two possible models include
Taylor dispersion and multiple transitions.  Taylor dispersion is
based on molecular diffusion
of particles in a flowing fluid (e.g., in a pipe) and is governed by an
ADE, to be discussed below. An identical formulation can be
obtained by considering particle movement in a random network
and applying the central limit theorem.
The extensive use of the
ADE in the hydrology literature is based essentially on the generic concept
of Taylor
dispersion and works well for relatively homogeneous systems.  The
particles are
assumed to be transported by the average flowing fluid in the medium while the
``diffusion'' is the dispersion due to local medium irregularities.
Larger scale effects
(e.g., permeability changes) are treated as perturbations of this model in
conventional stochastic treatments.

The prime interest in this work is in highly heterogeneous systems;
in these systems
contaminant motion can be envisioned as a migrating cloud of
particles, each of which
executes a series of steps or transitions between changes in velocity
{\bf v}. The
spatial extent of these transitions depends on the criterion used to
define changes in
{\bf v}. The classical approach is to consider the system
divided into representative elementary volumes (REVs) and determine an
{ \it average} {\bf
v} and dispersion {\bf D} in each REV. In our approach we dispense with the
REV idea, because averages can be unreliable in a system of very wide
fluctuations about the mean value. The change of concentration
$\Delta C$ at each
position in a  time increment $\Delta t$ is $\Delta t\times$(the net
particle flux). The
effective  volume contributing the net particle flux in $\Delta t$
can vary considerably
at different positions in the system. Thus the length scale over
which $\Delta C$
varies slowly in space can change considerably over the system. If one fixes a
sampling volume at each position, it is important to retain the full
distribution (not an
average) of the transition times (determined with a physical model)
of flux contributing
to $\Delta C$. If this distribution is retained, then in our approach
one can still use
the limit of a spatial  continuum (as shown below).

The distribution of transition times, $\psi(t)$, can be
determined in principle from an analysis of the
streamtubes of the flow field and contains the subtle features that can produce
non-Fickian behavior. The physical features necessary for non-Fickian
transport are the
existence of a wide range of transition times (causing large
differences in the flow
paths of migrating particles) and sufficient encounter with
statistically rare, but
rate-limiting slow transitions (e.g., low velocity regions) [{\it Berkowitz and
Scher}, 1995]. These general ideas will be developed schematically in
the next sections.

\subsection*{2.2. Single Realization Transport Equation}

For our point of departure we need a transport equation framework that
can enumerate all these possible paths and encompass the motion from
continuous to discrete over a range of spatial and temporal scales,
for any given realization of the domain.
An excellent candidate is the ``Master Equation''
[{\it Oppenheim et al.}, 1977; {\it Shlesinger}, 1996]

\be
\label{eq:master}
{ { {\partial C \left( {\bf s}, t \right) } \over
{ \partial t }} = -  \sum_{\bf s'} w ( {\bf s ' , s}) C ( {\bf s} , t)
+ \sum_{\bf s'}
w ( {\bf s  , s'}) C ({\bf s}',t) }
\ene
for $C({\bf s},t)$, the particle concentration at point {\bf s} and time
$t$, where $w({\bf s, s'})$ is  the transition rate from {\bf s}$'$
to {\bf s} (the
dimension of ${\Sigma}_{\bf s} w$ is reciprocal time). The transition rates
describe the
effects of the velocity field on the particle motion; the determination of
$w({\bf s, s'})$
involves a detailed knowledge of the system.
We assume the average effective range of $w({\bf s, s'})$
is a finite distance.
The Master Equation has been applied in the context of
electron hopping in random systems [e.g., {\it Klafter and
Silbey}, 1980a], and
is discussed widely in the physics and chemistry literature.

The transport equation in
(\ref{eq:master}) does not separate the effects
of the varying velocity field into an advective and
dispersive
part of the motion; this separation is an approximation based on the
assumption of
relatively homogeneous regions in which $C({\bf s},t)$ will be slowly
varying over a finite length scale (the range of transition rates),
\begin{widetext}
\begin{eqnarray}
\label{eq:expanC}
C ({\bf s '}, t) \approx C ( {\bf s}, t) + ( {\bf s ' -s}) \cdot \nabla C
( {\bf s } , t)
    + {\textstyle {1 \over 2}} ( {\bf s ' -s}) ( {\bf s ' -s}) :
\nabla \nabla C ( {\bf s}, t)
\end{eqnarray}
(with the dyadic symbol : denoting a tensor product).
Substituting
(\ref{eq:expanC})
into (\ref{eq:master}) leads to a continuum description
(i.e., local diffusion in a
pressure field $ \pi$({\bf s})) and a partial differential equation (pde),
for a single realization of the domain:
\begin{eqnarray}
\label{eq:pde1}
      { { \partial C ( {\bf s} , t ) } \over { \partial t }} &=&
\sum_{\bf s'}
( w ( {\bf s , s'}) - w ( {\bf s ', s})) C ( {\bf s} , t) +
\sum_{\bf s'} w ( {\bf s , s '}) ( {\bf s '- s}) \cdot \nabla C ({\bf
s} , t )
\nonumber \\
  &&+\sum_{\bf s'}  w ( {\bf s , s'}) {\textstyle { 1 \over 2}} ( {\bf s '- s})
( {\bf s '- s}) : \nabla \nabla C ( {\bf s} , t).
\end{eqnarray}
\end{widetext}

We note that (\ref{eq:pde1}) is close to the form of an ADE with the
exception of the
term proportional to $C({\bf s},t)$. This term is present due to the
asymmetry of the
transition rates (due to the bias of the pressure field) and/or the
non-stationary
medium (due to the explicit position dependence of the rates -- cf.
(\ref{eq:transrate})). It makes a contribution to the final form of the pde for
diffusion in a force field. If the system is stationary this term
vanishes (as we show
below) and thus reduces to the form of an ADE. One can already observe in
(\ref{eq:pde1}) generalized velocity and dispersion coefficients (in
terms of $w ( {\bf
s , s '})$); however we have not yet separated out the effects of the
flow field and
determined transport coefficients. In order to fully determine the
final pde and
separate the advection and diffusion contributions, we must specify
the $w({\bf s},{\bf
s}')$ in terms of $\pi({\bf s})$, the pressure field.

A general form for a non-stationary medium is
\be
\label{eq:transrate}
      w ({\bf s}, {\bf s}') \equiv W \left( | {\bf s}' - {\bf s} | ; {\bf s}'
      \right) \Omega ( \pi ({\bf s}') - \pi ({\bf s}))
\ene
where the asymmetry in the rates is due to $\pi({\bf s}') - \pi({\bf s})$, the
pressure difference at ${\bf s}'$ and ${\bf s}$, and the explicit
dependence of the
overall rate W on location ($\Omega$ is a function of the pressure difference
only). We specify the $\Omega$-function, so that (\ref{eq:transrate})
is written as
\begin{eqnarray}
\nonumber
      w ({\bf s}, {\bf s}') &\equiv& W \left( | {\bf s}' - {\bf s} | ;
       {\bf s}' \right) \Omega ( \pi ({\bf s}') - \pi ({\bf s}))\\
       &\approx&  F ( |{\bf s}' - {\bf s} | ; {\bf s}' )
       \left[ \lambda + {\textstyle {1 \over 2}} ( \pi ({\bf
         s}') - \pi ({\bf s})) \right]
\label{eq:translin}
\end{eqnarray}
where in (\ref{eq:translin}) non-linear terms in the pressure difference
have been neglected (i.e., terms proportional to $(\nabla \pi)^2$)
and a contribution to
the transition rates is retained even for vanishing pressure difference. The
significance of the latter step can be seen by realizing that $F (
\pi ({\bf s}') -
\pi({\bf s}))$ is a simple advection contribution (with a
permeability proportional to
$F$) and the term
$F\lambda$ is proportional to a local diffusion contribution to the
rates. The $\lambda$
term retains the scattering effects of the medium (i.e., the transfers between
``streamtubes'') even in the limit of very small local pressure
differences. It is also
closely associated with the effect of ``local'' dispersion.

We now also assume $F ( |{\bf s}' - {\bf s} | ; {\bf s}' )$ will be
slowly varying over
some finite length scale. We expand in a Taylor series to second
order in ${\bf s}' -
{\bf s}$,
\begin{eqnarray}
\nonumber
     F ( | {\bf s} ' - {\bf s} | ; {\bf s} ' ) &\approx& F ( | {\bf s}' -
      {\bf s} | ; {\bf s} ) + ( {\bf s}' - {\bf s} ) \cdot \nabla F\\
     &&+ {\textstyle {1 \over 2}} ( {\bf s}' - {\bf s}) ({\bf s}' - {\bf s}) :
      \nabla \nabla F .
\label{eq:expanF}
\end{eqnarray}
In (\ref{eq:expanF}), the gradient operates on the second argument, {\bf s}$'$.
Combining (\ref{eq:translin}) and (\ref{eq:expanF}), and substituting
into the first term on the right side of (\ref{eq:pde1}), we have
\begin{widetext}
\begin{eqnarray}
\nonumber
w ( {\bf s}, {\bf s}') - w ( {\bf s}',{\bf s}) &\approx& F ( | {\bf s}'
      -{\bf s} | ; {\bf s}') \left[ \lambda + {\textstyle {1 \over 2}}
      ( \pi ({\bf s}') - \pi
      ({\bf s})) \right]
     -F (|{\bf s}' - {\bf s}|; {\bf s})
       \left[ \lambda -  {\textstyle {1 \over 2}}
      ( \pi ({\bf s}') - \pi ({\bf s})) \right]\\
&\approx& F (| {\bf s}' - {\bf s}|;{\bf s}) (\pi ({\bf s}') - \pi ({\bf
s}))
\nonumber \\
&&+ \left[( {\bf s' - s}) \cdot \nabla F + {\textstyle {1\over 2}}
({\bf s'-s})
({\bf s' - s}) : \nabla \nabla F \right] \times
       \left[\lambda + {\textstyle {1\over 2}}
      (\pi ( {\bf s}') - \pi ({\bf s})) \right] .
\label{eq:expant1}
\end{eqnarray}
\end{widetext}
Now using a similar expansion for the pressure difference, we have
\be
\label{eq:expanpi}
      \pi ({\bf s}') - \pi ({\bf s}) \approx ({\bf s'-s}) \cdot \nabla
      \pi ({\bf s}) + {\textstyle {1\over 2}} ({\bf s ' - s})
      ({\bf s ' -s}) : \nabla \nabla \pi ({\bf s}) .
\ene
Substituting (\ref{eq:expanpi}) into (\ref{eq:expant1}) and using
\be
\label{eq:sumF0}
      \sum_{{\bf s}'} F (| {\bf s}'-{\bf s} | ; {\bf s} )
      ({\bf s}'-{\bf s}) = 0
\ene
because $F$ is an even function of the vector difference,
we obtain for the expression in (\ref{eq:expant1}), summed over s$'$,
\be
\label{eq:term1}
      \nabla \cdot  \left( {{ {\bf D}({\bf s})} \over { \lambda}}
      \nabla \pi ({\bf s}) \right)  +
      \nabla \cdot \nabla {\bf D}({\bf s})
\ene
where the dispersion tensor is defined as
\be
\label{eq:diffdef}
      {\bf D}({\bf s}) \equiv {\textstyle {1\over2}} \sum_{{\bf s}'} F (|
      {\bf s}'-{\bf s}|;{\bf s}) ({\bf s}' -{\bf s}) ({\bf s}'-{\bf s})
      \lambda .
\ene
We insert (\ref{eq:term1}) into (\ref{eq:pde1}) and use
(\ref{eq:transrate})--(\ref{eq:expanF}), (\ref{eq:expanpi}),
(\ref{eq:sumF0}) (cf.
Appendix A) to obtain
\be
\label{eq:pdef}
      {{\partial C ({\bf s},t)} \over {\partial t}} = \nabla \cdot
      \left[ {{{\bf D}({\bf s})} \over {\lambda}} \nabla \pi ({\bf s})  C({\bf
s},t) +
\nabla
      ({\bf D}({\bf s}) C({\bf s}, t))\right] .
\ene

The form of (\ref{eq:pdef}) is a continuity equation -- the time derivative
of the
concentration is equal to the divergence of the total concentration
flux, the sum of the
diffusive concentration flux and the advective concentration
flux -- with an effective permeability of
\be
\label{eq:permdef}
      {\bf k} (\bf s) \equiv {{ {\bf D}(\bf s) } \over { \lambda }} .
\ene
Equation (\ref{eq:pdef}) is a generalization to a non-stationary medium of the
well-known Smoluchowski equation [{\it Chandrasekhar}, 1943] which is the
basis for
describing diffusion in a
force field. In our case the force field is $\nabla \pi ({\bf s})$. In the
case of
electron transfer in a potential field the $\lambda$ in (\ref{eq:permdef})
can be shown
to be $\kappa$T (where T is the temperature and $\kappa$ is
Boltzmann's constant) and the relation in (\ref{eq:permdef}) is the Einstein
relation between mobility and diffusion.
We use a convention that a product between a tensor
{\bf T} and a vector {\bf V} is {\bf T}{\bf V} yielding a vector. In our
case, the
vector ${\bf v}(\bf s) = - {\bf k} (\bf s) \nabla \pi ({\bf s})$ is the
velocity
field and for an incompressible fluid $\nabla \cdot {\bf v}({\bf s}) = 0$. The
only term remaining in (\ref{eq:pdef}) proportional to $C({\bf s},t)$ is $
\nabla \cdot
\nabla {\bf D}({\bf s}) C({\bf s},t)$. The final form for the pde for
an incompressible fluid is
\be
\label{eq:pdeff}
      {{\partial C ({\bf s},t)} \over {\partial t}} = - {\bf v}({\bf s})
\cdot\nabla C({\bf
s},t) + \nabla \cdot\nabla ({\bf D}({\bf s}) C({\bf s}, t)) .
\ene

Equation (\ref{eq:pdeff}) is a generalization of the ADE.
While many simplifications of the ADE are based on
(\ref{eq:pdeff}) with ${\bf D}({\bf s}) ={\bf D}$ (i.e., a constant), the usual
(``general") form of the ADE includes a {\bf s}-dependent {\bf D} in
(\ref{eq:pdeff})
but with the second term replaced by $\nabla \cdot({\bf D}({\bf s})
\nabla C({\bf
s},t))$. Thus (\ref{eq:pdeff}) differs from this usual form of the
ADE by the addition of
two terms: $ \nabla \cdot \nabla {\bf D}({\bf
s}) C({\bf s},t)$ and $\nabla {\bf D}({\bf s}) \cdot \nabla C $.
The form of (\ref{eq:pdeff}) is the same as postulated by
{\it Kinzelbach} [1986], based on the Ito process.

The difference in the
general form of the ADE can be traced to starting the derivation with
the pressure field
$\pi({\bf s})$ and not with $\nabla\pi({\bf s})$, i.e., the expansion
(\ref{eq:expanpi}) is treated on the same basis as the other
expansions (\ref{eq:expanC})
and (\ref{eq:expanF}). Hence, starting with the Master equation
(\ref{eq:master}) and
using a general expression for the transfer rates we obtain, for a
specific heterogeneous
medium, in a continuum limit (slowly varying $C({\bf s},t)$ and
$w({\bf s},{\bf s}') )$
the generalized equation for diffusion in a force field
(Smoluchowski) which for
irrotational flow is a generalized ADE. We assert that for a
non-stationary medium, i.e., {\bf s}-dependent {\bf v} and {\bf D},
(\ref{eq:pdeff})
should be the starting point for numerical calculations. The main numerical
differences between this equation and the usual ADE (with {\bf
D}({\bf s}))  should
arise in ``boundary'' regions of more spatially varying {\bf D}({\bf s}).
The importance of accounting for {\bf D}({\bf s}) has been
demonstrated by, e.g., {\it
Labolle et al.} [1996].

We will show that the ``standard'' ADE emerges as the
continuum limit of the ensemble averaged Master equation (the term
proportional to
$C({\bf s},t)$ vanishes for stationary transition rates).
In general, the continuum limit presents difficulties in regions of increased
heterogeneity, such as tightly interspersed permeability layers. The
concentration
$C({\bf s},t)$ will not necessarily vary slowly on the  same length
scale throughout the
system. The point average of {\bf v} and {\bf D} can be very
sensitive to small changes
in the local volume used to determine the average. Conversely, if one
fixes the volume
to a practical pixel size (e.g., 10 m$^3$) the use of a local average
{\bf v} and {\bf
D} in each volume can be quite limited, i.e., the spreading effects
of unresolved
residual heterogeneities are suppressed [e.g., {\it Dagan}, 1997]. We
will return to this
issue in a broader context in section 4. It essentially involves the degrees of
uncertainty and its associated spatial scales. We start, at first,
with an ensemble
average of the entire medium and discuss the role of this approach in
the broader
context.

\subsection*{2.3. Ensemble Average Transport Equation}

We resume our examination of the Master Equation approach, i.e., before
assuming any continuum limit.  The ensemble average of (\ref{eq:master})
can be shown [{\it Klafter and Silbey}, 1980b] to be of the form
\begin{widetext}
\begin{eqnarray}
\label{eq:GME}
     {{ \partial P ({\bf s},t)} \over {\partial t}} = -
      \sum_{\bf s'} \int_0^t \phi ({\bf s}'-{\bf s}, t-t') P({\bf s},t') dt'
      + \sum_{\bf s'} \int_0^t \phi ({\bf s}-{\bf s}', t-t') P
      ({\bf s}', t') dt'
\end{eqnarray}
\end{widetext}
where $P({\bf s},t)$ is the normalized concentration, and $\phi({\bf s},t)$ is
defined below in (\ref{eq:phipsi}).  The form of (\ref{eq:GME}) is a
``Generalized
Master Equation" (GME) which, in contrast to (\ref{eq:master}), is
non-local in time and
the transition rates are stationary (i.e., depend only on the
difference {\bf s} -- {\bf
s}$'$) and time-dependent. This equation describes a semi-Markovian
process (Markovian in
space, but not in time), which accounts for the time correlations (or
``memory'') in
particle transitions.

It is straightforward to show [{\it Kenkre et al.}, 1973; {\it
Shlesinger}, 1974], using
the Laplace transform, that the GME is completely equivalent to a
continuous time random
walk (CTRW)
\be
\label{eq:CTRW}
      R({\bf s},t) = \sum_{\bf s'} \int_0^t \psi ({\bf s}-{\bf s}', t-t')
      R ({\bf s}', t') dt'
\ene
where $R({\bf s},t)$ is the probability per time for a walker to just arrive at
site  {\bf s} at time $t$, and $\psi({\bf s},t)$ is the probability rate for a
displacement {\bf s} with a difference of arrival times of $t$. The initial
condition for $R({\bf s},t)$ is $\delta_{{\bf s},0} \delta (t-0^+)$, which
can be appended to (\ref{eq:CTRW}).  The correspondence between
(\ref{eq:GME}) and
(\ref{eq:CTRW}) is
\be
\label{eq:PRrel}
      P({\bf s},t) = \int_0^t \Psi (t-t') R ({\bf s},t') dt'
\ene
where
\be
\label{eq:Psit}
      \Psi (t) = 1 - \int_0^t \psi (t') dt'
\ene
is the probability for a walker to remain on a site,
\be
\label{eq:psit}
      \psi (t) \equiv \sum_{\bf s} \psi ({\bf s},t)
\ene
and
\be
\label{eq:phipsi}
      \tilde{\phi} ({\bf s},u) = {{ u \tilde{\psi} ({\bf s}, u) } \over
      {1 - \tilde{\psi} (u)}}
\ene
where the Laplace transform ($\cal L$) of a function $f(t)$ is
denoted by $\tilde{f}(u)$.

Equations (\ref{eq:CTRW})--(\ref{eq:psit}) are in the form of a
convolution in space and
time and can therefore be solved by use of Fourier and Laplace
transforms [{\em Scher
and Lax}, 1973]. The general solution is
\begin{equation}
\label{ctrw1}
{\cal P}({\bf k},u)=\frac{1-\tilde{\psi}(u)}{u}\frac{1}{1-\Lambda({\bf k}, u)}
\end{equation}
where ${\cal P}({\bf k},u)$, $\Lambda ({\bf k}, u)$  are the Fourier transforms
(${\cal F}$) of $\tilde{P}({\bf s},u)$, $\tilde{\psi}({\bf s},u)$,
respectively.

The CTRW accounts naturally for the cumulative effects of a sequence of
transitions. The challenge is to map the important aspects of the
particle motion in the
medium onto a $\psi({\bf s},t)$. The identification of $\psi({\bf
s},t)$ lies at the
heart of the CTRW formulation. The CTRW approach allows a
determination of the evolution
of the particle distribution (plume), $P({\bf s},t)$, for a general
$\psi({\bf s},t)$; there is no a priori need to consider only the moments of
$P({\bf s},t)$. As we discuss below, a $\psi({\bf s},t)$ with a power law
(\ref{eq:asympsi}) for large time leads to the description of
anomalous transport (e.g.,
non-Fickian plumes). Once $\psi({\bf s},t)$ is defined one needs to calculate
$\Lambda({\bf k}, u)$ and then determine the propagator $P({\bf
s},t)$ by inverting the
Fourier and Laplace transform of (\ref{ctrw1}). The latter can be
quite challenging.

As shown previously the separation between advection and dispersion
occurs in the
continuum (diffusion) limit. In an ensemble averaged system this
limit leads to an ADE [{\it Berkowitz and Scher}, 2001]. For clarity
and convenience, we reproduce the argument here.
The first step is to make a series expansion of
$P({\bf s},t)$ similar to (\ref{eq:expanC}); inserting this
into (\ref{eq:GME}) yields
\begin{widetext}
\begin{eqnarray}
\label{eq:nonlocpde}
{{ \partial P ({\bf s},t)} \over {\partial t}} =
      \sum_{\bf s'}\int_0^t dt'  [\phi ({\bf s} - {\bf s}',t-t')
      ({\bf s}'-{\bf s})
      \cdot \nabla P({\bf s},t')
     + \phi ({\bf s} -{\bf s}',t-t') \textstyle{1 \over 2}
      ({\bf s}'-{\bf s}) ({\bf s}'- {\bf s}) :
      \nabla \nabla P({\bf s},t')] .
\end{eqnarray}
We write (\ref{eq:nonlocpde}) in a more compact form
\begin{eqnarray}
\label{eq:sumpde}
     {{ \partial P ({\bf s},t)} \over {\partial t}} =
      \int_0^t dt'[-{\bf v}_{\psi}(t-t')
      \cdot \nabla P({\bf s},t') +
     {\bf \Phi}_{\psi}(t-t'):
      \nabla \nabla P({\bf s},t')]
\end{eqnarray}
\begin{equation}
\label{eq:vsumpde}
{\bf v}_{\psi}(t)\equiv\sum_{\bf s} \phi ({\bf s}, t) {\bf s} \\
\end{equation}
\begin{equation}
\label{eq:Phisumpde}
{\bf \Phi}_{\psi}(t)\equiv\sum_{\bf s}\phi ({\bf s} ,t) \textstyle{1 \over 2}
{\bf s}{\bf s}
\end{equation}
\end{widetext}
Note the sum (over ${\bf s'}$) in (\ref{eq:nonlocpde}) is independent
of ${\bf s}$ in a
stationary system; hence we shift the summation variable to obtain
(\ref{eq:vsumpde})-(\ref{eq:Phisumpde}). This particular formulation
is convenient
because, in (\ref{eq:sumpde}), we can define terms that are familiar
in the context of
traditional modeling: the ``effective velocity''
${\bf v}_{\psi}$ and the ``dispersion tensor''
${\bf \Phi}_{\psi}$. Note, however, that both of these terms
are time-dependent, and most significantly, depend fundamentally
on $\psi({\bf s},t)$. This equation has the form of an ADE
generalized to non-local time
responses as a result of the ensemble average.

The next step is a crucial one in
distinguishing between normal and anomalous
transport. If $\psi({\bf s},t)$ has both a finite first and second moment
in $t$ the
transport is normal and one can expand $\tilde{\psi}({\bf s},u)$
as [{\it Scher and Montroll}, 1975]
\begin{eqnarray}
\label{eq:psiexpan1}
      \tilde{\psi} ({\bf s},u) \cong p_1 ({\bf s}) - p_2 ({\bf s}) u
                + p_3 ({\bf s}) u^2 + ...\; \; \;
        \nonumber \\ \mbox{and} \; \; \; \tilde{\psi} (u) = \sum_{\bf s}
\tilde{\psi}
       ({\bf s},u) \cong 1 - \bar{t} u + du^2 + ...
\end{eqnarray}
with $\sum_{\bf s} p_1 ({\bf s}) = 1$, the normalization of $\psi
({\bf s},t)$, and
$\sum_{\bf s} p_2 ({\bf s}) \equiv \bar{t}$ and
$\sum_{\bf s} p_3 ({\bf s}) \equiv d$,
the first and second temporal moments of $\psi(t)$, respectively.
Note that small $u$ corresponds to large time in Laplace space. The
functions $p_i({\bf
s})$ are asymmetric  due to the bias in the velocity field; $p_1({\bf
s})$ is the
probability to make a step of displacement {\bf s}. One now inserts
(\ref{eq:psiexpan1})
into (\ref{eq:phipsi}) and expands in a power series of $u$.  The
leading term is
independent of $u$, which we retain. The correction to this leading
term is proportional
to $u$ and is small. Substituting this expression into the Laplace transform of
(\ref{eq:sumpde})-(\ref{eq:Phisumpde}), which is
(\ref{eq:Lnlpde})-(\ref{eq:PhiLnlpde})
(cf. below), and taking the inverse Laplace transform of the result,
yields the ADE
\be
\label{eq:ADE}
      {{ \partial P ({\bf s},t)} \over {\partial t}} = -{\bf v} \cdot
      \nabla P ({\bf s},t) + {\bf D}: \nabla \nabla P ({\bf s},t)
\ene
where the effective velocity {\bf v} is equal to the first spatial moment of
$p_1({\bf s}), \bar{\bf s}$,  the mean displacement for a single
transition, divided
by the mean transition time  $\bar{t}$, and the dispersion tensor ${\bf D}
\equiv
D_{ij}$ is the second spatial moment divided by $\bar{t}$, which can be
written as
\be
\label{eq:vdef}
      {\bf v} = \sum_{\bf s} p_1 ({\bf s}) {\bf s}/\bar{t}\equiv \bar{\bf
      s} / \bar{t}
\ene

\be
\label{eq:Ddef}
      D_{ij} = v \textstyle{1 \over 2} \sum_{\bf s} p_1 ({\bf s}) s_i
s_j / \bar{s}
\ene
where $v = |{\bf v}|$ and $\bar{s} = |\bar{\bf s} |$.
If we retain the term proportional to $u$ when inserting
(\ref{eq:psiexpan1}) into (\ref{eq:phipsi}), we obtain terms with
both spatial and temporal derivatives of $P({\bf s},t)$.

Thus, our underlying physical picture of advective-driven 
dispersion reduces to the familiar ADE when
one can assume smooth spatial variation of $P({\bf s},t)$ and 
finite first and second temporal moments of $\psi({\bf s},t)$.

\subsection*{2.4. Non-Fickian Dispersion}

When the $\psi({\bf s},t)$ has a power law (algebraic tail)
dependence on time at large $t$, i.e.,
\be
\label{eq:asympsi}
\psi({\bf s},t) \sim t^{-1-\beta}
\ene
the first and second temporal moments do not exist for $0 < \beta < 1$,
while the second temporal moment does not exist for $1 < \beta < 2$.
The dependence of
$\psi({\bf
s},t)$ in (\ref{eq:asympsi}) is a manifestation of a wide distribution of
event times
as encountered in highly heterogeneous media. The relation between the
power law
behavior (\ref{eq:asympsi}) and non-Fickian (anomalous)
transport has been well documented [e.g., {\it Scher and Montroll}, 1975;
{\it Berkowitz and Scher}, 2001]. We sketch the key points of that
relationship:
The form of $\psi({\bf s},t)$ at large time determines the time dependence
of the mean position $\bar{\ell}(t)$ and standard deviation
$\bar{\sigma}(t)$ of
$P({\bf s},t)$. In the presence of a pressure gradient (or ``bias''), and for
(\ref{eq:asympsi}), it can be shown [{\it  Scher and Montroll}, 1975;
{\it Shlesinger}, 1974]
for $0 < \beta < 1$ that
\be
\label{eq:lmomt}
   \bar{\ell}(t) \sim t^\beta
\ene
\be
\label{eq:stdmomt}
    \bar{\sigma}(t) \sim t^\beta
\ene
while for $1 < \beta < 2$
\be
\label{eq:lmomtB}
   \bar{\ell}(t) \sim t
\ene
\be
\label{eq:stdmomtB}
    \bar{\sigma}(t) \sim t^{(3-\beta)/2} .
\ene
Moreover, it can be shown that Fickian-like transport arises when
$\beta > 2$ [e.g., {\it Margolin and Berkowitz}, 2000].

The unusual time dependence of $\bar{\ell}(t)$ and $\bar{\sigma}(t)$ in
(\ref{eq:lmomt})-(\ref{eq:stdmomtB}), resulting from the infinite temporal
moments of
$\psi({\bf s},t)$ (i.e., the conditions of the central
limit theorem are not fulfilled), is the hallmark
of the non-Fickian propagation of $P({\bf s},t)$.
  This behavior is in sharp contrast to Fickian models where,
$\bar{\ell}(t) \sim t$
and $\bar{\sigma}(t) \sim t^{1/2}$ (as an outcome of the central limit
theorem) and the
position of the peak of the distribution coincides with
$\bar{\ell}(t)$. Note that in Fickian transport,
$\bar{\ell}(t)/\bar{\sigma}(t) \sim
t^{1/2}$; an important distinguishing feature of anomalous transport is that
$\bar{\ell}(t)/\bar{\sigma}(t) \sim constant$ for $0 < \beta < 1$,
and $\bar{\ell}(t)/\bar{\sigma}(t) \sim t^{(\beta - 1)/2}$ for $1 < \beta < 2$.
The relative shapes of the anomalous
transport curves,
and the rate of advance of the peak, vary strongly as a function of
$\beta$. Thus the
parameter $\beta$ effectively quantifies the contaminant dispersion; this
parameter
is discussed in detail by, e.g., {\it Margolin and Berkowitz} [2000,
2002] and {\it Berkowitz and Scher} [2001].
Hence, the crucial consideration for the appearance of
non-Fickian dispersion in a specified scale of a heterogeneous
medium are the physical criteria for the power law (\ref{eq:asympsi}) and its
(time) range of applicability.
Non-Fickian transport that displays these characteristics has been
documented in several analyses of numerical simulations,
and laboratory and field data
[{\it Berkowitz and Scher}, 1998; {\it Hatano and Hatano}, 1998;
{\it Berkowitz et al.}, 2000; {\it Kosakowski et al.}, 2001].

The large time regime of $\psi({\bf s},t) $ corresponds to the small $u$
regime for
its Laplace transform and the expansion in $u$ (for
(\ref{eq:asympsi}))
is quite different
from (\ref{eq:psiexpan1}) [{\it Shlesinger}, 1974], i.e.,
\be
\label{eq:psiexpan2}
      \tilde{\psi} ({\bf s},u) \cong p'_1 ({\bf s})-p'_2 ({\bf s})u^\beta
+ ...
\ene
for $u \rightarrow 0$ for $0 < \beta < 1$.
Inserting (\ref{eq:psiexpan2}) into (\ref{eq:phipsi}), parallel
to the development following (\ref{eq:psiexpan1}), yields
a transport equation from
(\ref{eq:nonlocpde}) which
remains non-local in time and is not the ADE. Our development
[{\it Berkowitz and Scher}, 1995] of non-Fickian
transport has been based directly on (\ref{eq:GME}).
In other words, solutions for the full evolution of a tracer plume,
as well as for breakthrough curves (i.e., spatial and temporal
distributions of tracer) can be derived directly from
(\ref{eq:GME}) [e.g., {\it Scher and Montroll},
1975; {\it Berkowitz and Scher}, 1997, 1998].
A (fractional) pde form of the transport equation,
derived from (\ref{eq:nonlocpde}) and holding only for the
power law dependence (\ref{eq:asympsi}),
i.e., a special case of CTRW, is exhibited in
section 3.2. We observe also that
the $u \rightarrow 0$ expansion of $\tilde{\psi}({\bf s}, u)$
for $1 < \beta < 2$ is similar to (\ref{eq:psiexpan1}), but with
the $u^2$ term replaced by one proportional to $u^{\beta}$.
In this case the correction to the $u$-independent term
$p_1 ({\bf s}) / \bar{t}$ used in (\ref{eq:vdef}), (\ref{eq:Ddef})
is proportional to $u^{\beta - 1}$ and can be significant
(especially for $\beta \approx 1$).

Finally, we note that the general CTRW formalism
(i.e., not restricted to (\ref{eq:asympsi})) can
be used
to model a large number of physical processes.
For example, $\psi({\bf s},t)$ has been defined
for multiple trapping [e.g., {\it Scher et al.}, 1991; {\it Hatano and
Hatano}, 1998]
and as such can be used for multiple-rate models [{\it Haggerty and
Gorelick}, 1995]
and to quantify dispersion in stratified
formations [{\it Matheron and de Marsily}, 1980]. {\it
Zumofen et al.} [1991] have used the CTRW explicitly
to model the latter.

\section*{3. Fractional Differential Equations}
There is growing interest in the development and application of
fractional differential formulations of transport equations.
In particular, fractional differential equations of the diffusion,
diffusion--advection, and Fokker--Planck type have
been considered in stochastic modeling in physics
[e.g., {\it Hilfer}, 2000; {\it Metzler and Klafter}, 2000].
Here we consider fractional derivative equations (FDE) for
transport and show how they are special cases of the CTRW
equations developed in the previous section. We emphasize
that FDE are not different models from the CTRW; rather, they
are seen to emerge as asymptotic limit cases of the CTRW theory.

A word of caution: referring to a transport equation as ``fractional'' can
be with respect to the occurrence of fractional order
differentiation in time or space, or both. Moreover, a number of 
definitions for
fractional operators exist. Here, we concentrate on two possibilities:
the Riemann--Liouville fractional time derivative $_0D_t^{\beta}$
(for which we will employ the more suggestive notation 
$\partial^{\beta}/\partial
t^{\beta}$), and the Riesz spatial derivative $\nabla^{\mu}$ [{\em Oldham and
Spanier}, 1974; {\em Samko et al.}, 1993].

The development of FDE in both the time and space variables 
necessitates a more general
starting equation than (\ref{eq:nonlocpde}), which depends on the 
validity of the
expansion of $P({\bf s},t)$ similar to (\ref{eq:expanC}). We return 
to the general
solution (\ref{ctrw1}). In what follows, in order to obtain FDE's, we 
need the product
form $p({\bf s})\psi(t)$ for the $\psi ({\bf s},t)$ probability 
density function, which
assumes that the transition length and time are statistically 
independent quantities.
Furthermore we need the asymptotic form (\ref{eq:asympsi}) of 
$\psi(t)$ and/or $p({\bf
s})$ (cf. below). The indicated power-law decay for $0<\beta<1$ 
causes the divergence of
$\bar{t}$, the mean transition time (cf. section 2.4). Corresponding to
(\ref{eq:asympsi}) the Laplace transform of $\psi(t)$ is
\begin{equation}
\label{eq:asymltpsi}
\tilde{\psi}(u)\sim 1-(uc_t)^{\beta}
\end{equation}
which is (\ref{eq:psiexpan2}) summed over ${\bf s}$, where $c_t$ is
a dimensional constant determined by the physical model. Along the 
same line we consider
the power--law form $p({\bf s})\sim {c_s}^{\mu}/|{\bf 
s}|^{1+\mu},\quad 0<\mu<2$ for the
transition length, where $c_s$, is analogous to $c_t$, a dimensional 
constant. Similar
to $\psi(t)$, the first and second or second (spatial) moment(s) of 
$p({\bf s})$ are
infinite for, respectively, $0 < \mu\ < 1$ and $1 < \mu\ < 2$. The 
border case for $\mu
= 2$ is the Gaussian law $p({\bf s})\sim(4\pi{c_s}^2)^{-1}\exp(-{\bf 
s}^2/(4{c_s}))$. For
any symmetric L{\'e}vy stable law $p({\bf s})$, the asymptotic form 
of the Fourier
transform of $p({\bf s})$ is given by
\begin{equation}
\label{eq:asymftp}
\tilde{p}({\bf k})\sim 1-{c_s}^{\mu}|{\bf k}|^{\mu} \qquad \qquad  0<\mu\le 2.
\end{equation}
\subsection*{3.1 Time-FDE}
We concentrate on the case $0 < \beta < 1$ and $\mu=2$,
for which the spatial moments are finite, but the temporal moments 
are infinite.
We consider first the case with no spatial bias, $\bar{\ell}(t)=0$ (i.e., no
advective transport). Insertion of (\ref{eq:asymltpsi}) and the low wavenumber
expression $\tilde{p}({\bf k})\sim 1-{c_s}^2{\bf k}^2$ into 
(\ref{ctrw1}) leads to
\begin{equation}
\label{prop1}
\tilde{\cal P}({\bf k},u)=\frac{1}{u+K_{\beta}u^{1-\beta}{\bf k}^2}
\end{equation}
(dropping the cross term $(uc_t)^{\beta}{c_s}^2{\bf k}^2$)
where the anomalous diffusion constant is defined as $K_{\beta}\equiv
{c_s}^2/{{c_t}^{\beta}}$. The FDE is determined by multiplying (\ref{prop1})
by the denominator of the right side and rearranging to yield
\begin{equation}
\label{prefd}
u {\cal P}({\bf k},u)-1=-K_{\beta}{\bf k}^2u(u^{-\beta}{\cal P}({\bf k},u)),
\end{equation}
where the dimension of the generalized diffusion constant is
$[K_{\beta}]={\rm cm}^2{\rm sec}^{-\beta}$.
While the two terms on the left correspond to $\partial P({\bf 
s},t)/\partial t$
in $({\bf s},t)$ space, with the initial condition $P({\bf s}, 
0)=\delta({\bf s})$
(on both sides of (\ref{prefd}) the property ${\cal L}
\{dF(t)/dt\}=u\tilde{F}(u)-F(0)$ is utilized), the factor 
$u^{-\beta}$ on the right poses
the problem of finding the corresponding Laplace inversion. One of 
the definitive
responses goes back to Riemann and Liouville who extended the Cauchy 
multiple integral,
in order to define the fractional integral,
\begin{equation}
\label{rl}
\frac{\partial^{-\beta}}{\partial t^{-\beta}}
P({\bf s},t) \equiv\frac{1}{\Gamma(\beta)}
\int_0^t dt' \frac{P({\bf s},t')}{(t-t')^{1-\beta}}
\end{equation}
which possesses the important property
\begin{equation}
{\cal L}\left\{\, \frac{\partial^{-\beta}}{\partial t^{-\beta}}
P({\bf s},t)\right\}=u^{-\beta}\tilde{P}({\bf s},u).
\end{equation}
The definition (\ref{rl}) explicitly includes the initial value at time $t=0$.
Note that for a negative index, $\partial^{-\beta}/\partial t^{-\beta}$, the
Riemann--Liouville operator denotes fractional {\em integration} 
whereas for a positive
index, $\partial^{\beta}/\partial t^{\beta}$, we have fractional {\em
differentiation}. In our case fractional differentiation is established as the
succession of fractional integration and standard differentiation:
\begin{equation}
\frac{\partial^{1-\beta}}{\partial t^{1-\beta}}
P({\bf s},t)=\frac{\partial}{\partial t} \frac{\partial^{-\beta}}{
\partial t^{-\beta}} P({\bf s},t).
\end{equation}

With these definitions, we can now invert (\ref{prefd}), and obtain
the fractional diffusion equation
\begin{equation}
\label{fd}
\frac{\partial P}{\partial t}=K_{\beta} \frac{\partial^{1-\beta}}{
\partial t^{1-\beta}} \nabla^2P({\bf s},t).
\end{equation}
In the limit $\beta \to 1$ (\ref{fd}) reduces to the standard Brownian version.

The generalization to a fractional ADE for anomalous transport 
$(0<\beta<1)$, which
includes a spatial bias (advective transport), follows the same 
procedure as above
[{\it Compte}, 1997; {\it Compte et al.}, 1997; {\it Compte and 
C{\`a}ceres}, 1998; {\it
Metzler et al.}, 1998; {\it Metzler and Compte}, 2000],
\begin{equation}
\frac{\partial}{\partial t}P({\bf s},t)=\frac{\partial^{1-\beta}}{\partial
t^{1-\beta}}\left(-{\bf v}
_{\beta}\cdot\nabla+K_{\beta}\nabla^2\right)P({\bf s},t)
\label{fda}
\end{equation}
where ${\bf v}_{\beta}$ is the ``generalized drift velocity''.
Note that (\ref{fd}) and (\ref{fda}) involve fractional
differentiation in time on the spatial derivative terms of the equations.
These equations can be rewritten so they do not involve mixed 
derivatives, if desired
[{\em Metzler and Klafter}, 2000]. We stress that the form of 
(\ref{fd}) and (\ref{fda})
relies on using (\ref{eq:asymltpsi}), and that (\ref{fda}) is valid only for
$0<\beta<1$; it is modified significantly for $1<\beta<2$. We have 
thus shown that the
probability density $P({\bf s},t)$ described by the time--fractional 
ADE (\ref{fda}), is
equivalent to the large--time limit of the CTRW with a bias, with the 
asymptotic form of
$\psi (t)$ given by (\ref{eq:asympsi}) (or $\tilde \psi(u)$ given by
(\ref{eq:asymltpsi})).
For a specific class of $\psi (t)$ (which also fulfills the asymptotic
form (\ref{eq:asymltpsi})), the equivalence between
CTRW and FDE can be shown over the entire range of $t$
[{\em Hilfer and Anton}, 1995].

\subsection*{3.2. Space-FDE: L{\'e}vy Flights}
We now consider the opposite case of a transition time distribution 
with an existing
first moment, ${\beta}>1$, $\tilde{\psi}(u)\sim 1-uc_t$, and a 
transition length
distribution $p({\bf s})$ with a diverging second moment ($0<\mu<2$)
(${\cal F}\{p({\bf s})\}$ in (\ref{eq:asymftp})). This case can be 
shown to be a
Markovian process (in contrast to the semi-Markovian process 
discussed in section 2.3)
called a L{\'e}vy flight.

To avoid confusion, we stress that a L{\'e}vy flight refers to a 
random movement in
space, where the length of the transitions is considered at discrete steps, but
time is not involved. L{\'e}vy walks, on the other hand, attach a 
time ``penalty'', by
assigning a velocity to each transition in space. In the simplest 
case, this velocity is
constant; relaxation of this condition leads back to the more general 
CTRW formulation of
section 2.3 [{\em Klafter et al.}, 1987; {\em Shlesinger et al.}, 
1993].  In any case,
L{\'e}vy walks cannot be described in terms of simple fractional 
transport equations
[{\em Metzler}, 2000].

A L{\'e}vy flight is characterized through the Fourier--Laplace
transform [{\em Bouchaud and Georges}, 1990; {\em Compte}, 1996;
{\em Metzler and Klafter}, 2000]
\begin{equation}
\label{lflp}
\tilde{\cal P}({\bf k},u)=\frac{1}{u+K^{\mu}|{\bf k}|^{\mu}}
\end{equation}
from which, upon Fourier and Laplace inversion, the FDE
[{\it Compte}, 1996]
\begin{equation}
\label{lfd}
\frac{\partial}{\partial t}P({\bf s},t)=K^{\mu}\nabla^{\mu}P({\bf s},t)
\end{equation}
is inferred. The Riesz operator $\nabla^{\mu}$ is defined through 
[{\em Samko et al.},
1993]
\begin{equation}
{\cal F}\Big\{\nabla^{\mu}P({\bf s},t)\Big\}=
-|{\bf k}|^{\mu}{\cal P}({\bf k},t).
\end{equation}
Note that we use the definition $K^{\mu}\equiv {c_s}^{\mu}/c_t$
for the diffusion constant. From (\ref{lflp}), one recovers the characteristic
function
\begin{equation}
{\cal P}({\bf k},t)=\exp\left(-K^{\mu}t|{\bf k}|^{\mu}\right),
\end{equation}
which is the characteristic function of a centered and
symmetric L{\'e}vy distribution with the asymptotic power--law behavior
[{\it L{\'e}vy}, 1925, 1954; {\it Gnedenko and Kolmogorov}, 1954]
\begin{equation}
\label{lssol}
P({\bf s},t)\sim |{\bf s}|^{-1-\mu}.
\end{equation}
L{\'e}vy distributions are used to generate
L{\'e}vy flights [{\it Bouchaud and Georges}, 1990].
Accordingly, the second moment of a L{\'e}vy flight diverges:
\begin{equation}
\label{divmom}
\langle {\bf s}(t)^2\rangle=\infty .
\end{equation}
Observe that L{\'e}vy flights are characterized by a transition time 
distribution
$\psi(t)$ with a finite first moment; they are thus fundamentally 
different from those
processes underlying the time--fractional dispersion equation 
(\ref{fda}). As can be seen
both descriptions are included in the CTRW framework.

Including a bias into the transition distribution, one obtains
for an asymptotic form of $p({\bf s})$ the
L{\'e}vy flight fractional ADE [{\em Metzler et al.}, 1998]
\begin{equation}
\label{lfad}
\frac{\partial}{\partial t}P({\bf s},t)+{\bf v}\cdot\nabla P({\bf s},t)=
K^{\mu} \nabla^{\mu}P({\bf s},t)
\end{equation}
which exhibits Galilei symmetry, i.e., (\ref{lfad}) is solved by the
L{\'e}vy stable solution (\ref{lssol}), to be taken at the point ${\bf
s-v}t$. This means that the symmetric L{\'e}vy stable plume is
entirely shifted along the velocity vector ${\bf v}$, a situation which
strongly contrasts the growing skewness in the CTRW case for long-tailed
transition times.
Of course, this solution features the same divergence (\ref{divmom}) of the
second moment of the plume distribution. The first moment of (\ref{lfad})
exists for all $0<\mu<2$ and follows the usual Galilei symmetry expression
\begin{equation}
\langle {\bf s}(t)\rangle={\bf v}t.
\end{equation}

\subsection*{3.3. Applications}
As discussed above, although both time and space FDE forms are
special cases of the CTRW, and both represent generalizations of the
Fickian-based ADE, there are clear and critical distinctions between the
transport equations that result from these two
formulations. Here, we assess the L{\'e}vy flight description and argue that
its characteristics strongly limit its applicability to describing transport in
geological formations.

We consider the underlying physical picture of the L{\'e}vy flight,
as applied to tracer migration in geological formations:
a necessary condition for the L{\'e}vy flight
description is that the domain clearly contain ``streaks''
of high and low permeability, arranged so as to lead to particle transitions
of high and low velocity. In other words, the physical picture of a
L{\'e}vy flight requires an encounter with a wide range of lengths of
permeability streaks to obtain a non-Fickian distribution of particle 
transitions.
And yet, such non-Fickian distributions arise even without the 
presence of such a
permeability distribution, as clearly demonstrated by, e.g., {\it 
Silliman and Simpson}
[1987].

In addition, we observe that in mathematical terms, the first and 
second moments are
often used to characterize plume migration. These quantities describe the
spatio-temporal distribution of the tracer particles; the particles carry a
finite mass, and therefore have a finite velocity. As noted above, 
the L{\'e}vy flight
description leads to a diverging second moment of the migrating 
plume. Given that the
macrodispersion parameter is typically defined in terms of the second 
moment, this
divergence property cannot be ignored.  Moreover, we observe that 
through scaling
arguments  [{\em Jespersen et al.}, 1999], transport only undergoes a 
``superdiffusive''
(faster than linear) process; in the L{\'e}vy flight description, 
subdiffusive transport
can never occur.

With respect to the issue of a diverging second moment,
one might attempt to work with a finite number of sampled
tracer particles in a finite range, during a finite time
window; this leads to a truncated L{\'e}vy distribution with
finite moments. For truncated L{\'e}vy distributions it
is known that their scaling behaviors in time pertain up to relatively
large times [{\it  Mantegna and Stanley}, 1994, 1995].
The difficulty is that to account for the
temporal evolution of the particle cloud, the cutoffs would have to be
adjusted to the actual space volume explored by the tracer particles,
i.e., the cutoffs would themselves become time-dependent [{\em Jespersen
et al.}, 1999]. Put somewhat differently, the spatial-fractional
formulation is based on an assumed fractal, scale-free nature of
the transport process. Truncating
the distributions leads, by definition, to a scale-dependent process which
invalidates the use of simple fractional operators.

In contrast to the above arguments, the formulation given by,
e.g., (\ref{fda}), or, more generally, by
(\ref{eq:CTRW})-(\ref{eq:psit}), does not suffer
from these limitations or assumptions.
In realistic field situations, the distribution of particle
velocities is expected to vary widely on the order of
magnitude of typical spacing between sampling points.
Of course, the velocity distribution is bounded by some maximum
velocity. In the long time limit,
corresponding to the small $u$ limit that is of interest in our modeling,
the mean effect of this finite variation of velocities can be approximated
by a typical velocity. From this point of view, therefore, anomalies
in the plume and
the related moments should arise from temporal ``sticking'' processes
(i.e., low velocity particle transitions) which are taken
into consideration in the CTRW picture.
Depending on the range of $\beta$ (recall (\ref{eq:lmomt})--
(\ref{eq:stdmomtB})), both subdiffusive and superdiffusive behaviors
for plume spreading can be characterized. Moreover, explicit spatial
structure (well-defined conductivity features) can be incorporated
within the CTRW framework.

\section*{4. The use of CTRW-based ensemble averages
in non-stationary media: The
relation of field scales and uncertainty}

We return now to consider the issue, raised in the Introduction,
that the interplay between ensemble averaging
and spatial scales of non-stationary geological features
strongly affects efforts to model transport.
Broadly speaking, there exist two approaches to modeling
transport in large, field-scale formations.
In the first approach, the formation is treated as a
single domain, with heterogeneities characterized
and distributed according to a random field, with or
without correlation and/or anisotropy. Generally speaking,
these characterizations treat the domain as a stationary system,
although stochastic models that incorporate a deterministic
drift component (in the random field generator)
have been considered [e.g., {\it Li and McLaughlin}, 1995].
In the second approach, a physical picture of the domain
is constructed which includes explicitly
specified (prescribed or known) heterogeneities, so that
the resulting domains are non-stationary
[e.g., {\it LaBolle and Fogg}, 2001; {\it Koltermann and Gorelick}, 1996;
{\it Eggleston and Rojstaczer}, 1998; {\it Feehley et al.}, 2000].

While the study of ensemble-averaged (stationary) domains has given rise to a
sub-literature on stochastic methodologies and limiting behavior
(e.g., perturbation
techniques, macrodispersion) it has not yielded a practical numerical
scheme to deal
with the large majority of field sites. {\it Anderson} [1997] describes in
detail
heterogeneity  and trending structures evident in natural geological
formations,  and
argues convincingly for the need to use facies modeling (coupled with
geostatistical
techniques) and/or depositional simulation models. These models can provide
the underlying
hydraulic conductivity structure and flow field of non-stationary domains,
conditioned on
field measurements, and be integrated with predictive models of transport.

Within the framework of non-stationary domains, explicitly characterized
by structural trends, the question then arises as to how best to model
transport (or, more
precisely, how to  deal with the unresolved heterogeneities (residues)).
Clearly,
there is a critical interplay between length scales associated with the
trends and the
residues. This gives rise to the associated uncertainty in both the
measured/estimated hydraulic parameters and the measured/predicted
concentrations.
The generally accepted explanation for non-Fickian transport is that
heterogeneities which
cannot be  ignored are present at all scales. Therefore, accounting for
these residues is
a central consideration for the quantification of non-Fickian transport.

In efforts to combine non-stationarity with local-scale heterogeneity
and uncertainty, several recent studies have attempted to use
ADE-based modeling approaches
in conjunction with facies modeling [e.g., {\it Eggleston and Rojstaczer},
1998; {\it Feehley et al.}, 2000]. However, these
studies, which incorporated even highly discretized
systems (e.g., with block sizes of the order of 10 m$^3$ in
large aquifers), demonstrated an
inability to adequately capture the migration patterns;
these results suggests that unresolved heterogeneities also
exist at these relatively small scales. We note that non-Fickian
transport has been observed even in small-scale, relatively
homogeneous, laboratory-scale models [{\it Berkowitz et al.}, 2000].
Other related issues that have been considered recently focus
on the relative importance of diffusion and local-scale
dispersion and on
how to separate diffusive mass transfer processes from
slow particle velocities [e.g., {\it Harvey and Gorelick}, 2000;
{\it LaBolle and Fogg}, 2001].
These questions may be considered to be somewhat moot, especially
given that ``dispersion'' is an artifact of averaging in
mathematical formulations, while a definitive separation
between diffusion and very low velocity may be unnecessary.

At all of these smaller scales, i.e., within individual
facies or depositional structures, the CTRW-based transport
equations are highly effective.
We therefore suggest that the CTRW-based approach should be
used together with these facies and depositional models.
As is usually done, a numerical
model can be constructed which accounts explicitly for the heterogeneity
structure of a formation, and the usual methods to solve for the
flow field can be implemented. A CTRW-based transport equation can then
be applied, rather than the ADE, over the entire domain.
We observe that while the ADE (and the usual
definition of ``dispersion'') is simpler to apply than the CTRW-based
equation, the preceding discussion (both in this section and
the previous ones) demonstrate that it cannot and should not generally be
applied in realistic field situations.

In this context, we shall consider the use of a hybrid model: known
conductivity
structures are accounted for explicitly, and within each block (pixel
or voxel) of a
numerical model we use the CTRW to account for the residues.
Precluding the use of
$\psi({\bf s},t)$ with (spatial) L{\'e}vy forms, because the trends are
included explicitly in the
numerical model, we can start with (\ref{eq:sumpde}) as a basis for
our numerical
treatment. The methods developed with the use of the ADE, can be
carried out with the
Laplace transforms of (\ref{eq:sumpde})-(\ref{eq:Phisumpde}),
\begin{eqnarray}
\nonumber
u\tilde{ P} ({\bf s}, u) - P_{0}({\bf s}) &=& -\tilde{{\bf v}}_{\psi}(u)\cdot
\nabla
\tilde{ P} ({\bf s}, u)\\
&&+ \tilde{{\bf \Phi}}_{\psi}(u):\nabla \nabla \tilde{ P} ({\bf s}, u)
\label{eq:Lnlpde}
\end{eqnarray}
\begin{equation}
\label{eq:vLnlpde}
\tilde{{\bf v}}_{\psi}(u)= {{ u {\Sigma}_{\bf s} \tilde{\psi} ({\bf s}, u) {\bf
s}} \over
      {1 - \tilde{\psi} (u)}} \\
\end{equation}
\begin{equation}
\label{eq:PhiLnlpde}
\tilde{{\bf \Phi}}_{\psi}(u)= {{ u {\Sigma}_{\bf s} \tilde{\psi} ({\bf s}, u)
\textstyle{1 \over 2} {\bf s}{\bf s}} \over
      {1 - \tilde{\psi} (u)}}
\end{equation}
where $P_{0}({\bf s})$ is the initial condition.

The transport equation (\ref{eq:Lnlpde}) is very similar to the Laplace
transform of
the ADE, but with the important exception that
$\tilde{{\bf v}}_{\psi}$ and $\tilde{{\bf \Phi}}_{\psi}$
are $u$-dependent. A spatial grid can be employed to
numerically solve
(\ref{eq:Lnlpde}), exactly as can be done with the ADE
applied to a non-stationary system.
At each grid point, the velocity value determined from
the solution to the steady flow problem
is used in (\ref{eq:Lnlpde})-(\ref{eq:PhiLnlpde}), along with the
corresponding estimate of $\beta$, to change the
parameters
of $\tilde{\psi} ({\bf s}, u)$ and $\tilde{\psi} (u)$.

In this methodology the
interpretation of $\tilde{\psi} ({\bf s}, u)$ changes somewhat.
Instead of single
transitions, we consider
$\tilde{\psi} ({\bf s}, u)$ as
playing the role of accounting for the transition across an entire
element of the spatial grid. This interpretation has been
justified by {\it Margolin and Berkowitz} [2000].

If we insert (recall (\ref{eq:psiexpan2}))
\be
\label{eq:psivdep}
\tilde{\psi} (u) \cong 1-c_{\beta} u^{\beta}, \; \; \; \mathrm{for} \; \; \;
\; 0<\beta<1
\ene
into (\ref{eq:Lnlpde})-(\ref{eq:PhiLnlpde}), we
generate non-Fickian transport across each block element
(with $c_{\beta}$ proportional
to the velocity value at the grid point, divided by a characteristic
length, all raised to the $\beta$ power). The non-Fickian
behavior is due to the unresolved heterogeneities below the scale of the
spatial grid.
Estimates of $\beta$ and $c_\beta$ can be obtained for each
facies from a standard tracer breakthrough test and subsequent
comparison and fitting with analytical solutions (as done, e.g., in
{\it Berkowitz et al.} [2000] and {\it Kosakowski et al.} [2001]);
this procedure is exactly analogous to the usual
determination of the dispersivity parameter $\alpha$ in the ADE.

Using a more complete expression for $\tilde{\psi} ({\bf s}, u)$ we can
also evolve the
dynamics of the plume at very long time into a Gaussian (i.e., in a time
regime in which
$\psi ({\bf s}, t)$ possesses a finite first and second temporal moment).
The change in
$\tilde{\psi}
({\bf s}, u)$ across the boundaries can be handled by using suitable
averages similar to
the ADE-based numerical treatments. Hence one can numerically solve for
$\tilde{ P}
({\bf s}, u)$ at each grid point and obtain the normalized concentration $P
({\bf
s},t)$ by calculating  ${\cal L}^{-1}[\tilde{ P}({\bf s}, u)]$.
However, the inversion of
a Laplace transform can be challenging, and remains a key issue
for future research.

Finally, if we include pumping wells at some of
the grid points ${\bf s}_{p}$ (where
$\tilde{\psi} ({\bf s}_{p}, u)=0$, because the
particles enter the well but do not emerge),
then we can obtain
the accumulated concentration
directly from $\tilde{ P}({\bf s}_{p}, u\to 0)$.
In other words, $\tilde{ P}({\bf s}_{p}, 0) =
\int_0^{\infty} dt{ P}({\bf s}_{p}, t)$, and because
mass is conserved, each pumping well acts
as a sink extracting a fraction of the migrating
particles.

\section*{5. Summary and Conclusions}

The application of stochastic approaches
to quantification of transport in heterogeneous geological media
rests inevitably on the underlying conceptual picture
of dispersive mechanisms.
The fundamental significance of this picture was pointed out long ago.
As noted by {\it Bear} [1972], in his discussion of the work of
{\it Scheidegger} [1954, 1958], ``...the application of
the statistical approach requires...a choice of the type of statistics
to be employed, i.e., the probability of occurrence of events
during small time intervals within the chosen ensemble. This may
take the form of correlation functions between velocities at
different points or different times, or joint-probability
densities of the local velocity components of the particle as
functions of time and space or a probability of an elementary
particle displacment. The chosen correlation function determines
the type of dispersion equation derived.''

We have developed this early insight into a full, quantitative
theory where the joint probability density is the $\psi ({\bf s},t)$.
This joint spatial-temporal distribution allows us to account
for the behavior of migrating particles which can encounter
a wide range of velocity regions in heterogeneity lenses of
different spatial dimensions. This approach is in contrast
to most others which have, historically, emphasized spatial
formulations of transport equations, motivated by the
clear spatial heterogeneity of geological formations.

The overarching framework for our physical picture of
transport, and the assumptions
(as detailed above) on particle transitions, is the Master Equation.
This equation represents a general, yet highly applicable,
quantification of transport which recognizes the broad
spectrum of particle motions in space and time. We show,
under a general assumption of the form of $w({\bf s, s'})$,
that the Master Equation can be specialized in any single realization
of the geological domain to a generalized form of the ADE.

The ensemble average of the unrestricted
Master Equation leads to a Generalized Master Equation, which
is exactly equivalent to the CTRW. As a limiting
form, under highly restrictive conditions regarding
the character of the transport (and therefore of
the degree of structural heterogeneity), the conventional
ADE can be recovered from this formulation.

Aquifers are inherently heterogeneous over a wide range of scales,
and Fickian transport (embodied in the ADE) does not
generally occur on practical scales of interest.
We therefore suggest that the overwhelming
focus on defining ``effective'' dispersion, or
``macrodispersion'' coefficients, in Fickian or
pseudo-Fickian formulations
of the transport problem, is misplaced for field-scale problems.
The CTRW theory, which is the basis for our transport equation,
quantifies naturally the non-Fickian behavior observed
at laboratory and field scales, as well as in numerical
simulations. The essential character of the transport
can be embodied in an asymptotic form of the
$\psi ({\bf s},t)$, specifically by an exponent $\beta$.
This exponent, which can be determined from the velocity
distribution (based on solution of flow for a given
conductivity field) or from a tracer test, parameterizes
an entire class of non-Fickian plume evolutions, on scales
larger than the size of the heterogeneities.
Detailed discussions on the practical identification of
$\psi ({\bf s},t)$ and parameter values is given in
{\it Berkowitz and Scher} [2001], {\it Kosakowski et al.} [2000],
and {\it Berkowitz et al.} [2000, 2001].

We have also shown how fractional derivative formulations
of transport equations are special, asymptotic (limit)
cases, (\ref{eq:asympsi}) for $\psi ({\bf s},t)$,
of the CTRW theory.
Inserting this limiting form (\ref{eq:psiexpan2}) into the
Laplace transform of (\ref{eq:GME}), one arrives at the same step
necessarily encountered at the outset of the solution of the FDE.
Retention of the more general equation
(\ref{eq:GME}) has important advantages for a more complete modeling of the
transport process. The limiting forms characterized by the exponent
$\beta$ (which is the
fractional order of the  derivative in the FDE) apply for a certain
time range only.
Beyond this range, the $\psi ({\bf s},t)$ changes in a manner that
allows the plume to
eventually assume a Gaussian shape (defined by ``macrodispersion'')
as is reasonable
for most physical systems.

Finally, we consider how best to quantify contaminant transport in
non-stationary
geological  formations. We delineate a hybrid approach in which known
structural
properties are included explicitly, and  unresolved (unknown)
heterogeneities at smaller
scales  are accounted for within the CTRW theory. Practical
application of this approach
is achieved by replacing the usual ADE equation that is integrated
into numerical
simulation codes by a CTRW-based transport equation. This transport
model can be
integrated with existing  numerical modeling techniques to determine
the underlying
flow field.

We are currently focusing efforts on implementation of the solution
technique suggested
here, as well as on deriving analytical solutions for CTRW-based
transport equations for
forms of $\psi({\bf s},{\it t})$ generalized in both space and time.

\begin{center}
{\bf Appendix A}
\end{center}

We showed how the use of (\ref{eq:transrate})--(\ref{eq:expanpi})
leads to the expression (\ref{eq:term1}) for the first
term of the right side of (\ref{eq:pde1}).
We outline the derivation here for the second
and third terms of the right side of (\ref{eq:pde1}),
using these same equations.
We have for the second term
\begin{widetext}
\begin{eqnarray}
\nonumber
\sum_{\bf s'} w ({\bf s},{\bf s}') ({\bf s}'-{\bf s}) \cdot \nabla C
      ({\bf s},t)
   &\approx& \sum_{\bf s'} (F (|{\bf s}'-{\bf s}|;{\bf s}) + ({\bf s}'-{\bf s})
\cdot
      \nabla F) \left[ \lambda + \textstyle{1\over2} ({\bf s}'-{\bf s})
      \nabla \pi) \right] ({\bf s}'-{\bf s})\cdot
      \nabla C({\bf s},t) \nonumber \\
   &\approx& \sum_{\bf s'} \lambda F (|{\bf s}'-{\bf s}|;{\bf s}) \textstyle{1
\over 2}
      ({\bf s}'-{\bf s})
       {\displaystyle {{ \nabla \pi} \over {\lambda }} }
      ({\bf s}'-{\bf s}) \cdot  \nabla C ({\bf s},t)\nonumber\\
    &&+\sum_{\bf s'} ( {\bf s ' -s})\cdot
      \nabla F  \lambda ( {\bf s ' -s})\cdot \nabla C ({\bf s},t)\nonumber\\
  &=& {\bf D}({\bf s}) {{ \nabla \pi} \over {\lambda }} \nabla C ({\bf
s},t) + 2 \nabla
      {\bf D}({\bf s}) \nabla C ({\bf s},t)
\end{eqnarray}
and for the third term,
\begin{eqnarray}
\sum_{\bf s'} w({\bf s},{\bf s}') \textstyle{1\over2} ({\bf
s}'-{\bf s})({\bf
s}'-{\bf s}): \nabla \nabla C ({\bf s},t)
\approx \sum_{\bf s'} F(|{\bf s}'-{\bf
s}|;{\bf s}) \lambda \textstyle{1\over2} ({\bf s}'-{\bf s})({\bf s}'-
{\bf s}) :
\nabla \nabla C({\bf s},t)
     = {\bf D}({\bf s}) : \nabla \nabla C({\bf s},t)
\end{eqnarray}
\end{widetext}

We add the results of (A1), (A2) and (\ref{eq:term1}) to obtain
(\ref{eq:pdef}).
\section*{Acknowledgments}
The authors thank Marco Dentz and Gennady Margolin for useful
discussions, and two anonymous reviewers for constructive
comments. BB thanks the European Commission (Contract No.
EVK1-CT-2000-00062) for financial support.
RM acknowledges financial support from the Deutsche
Forschungsgemeinschaft within the Emmy Noether program.

\section*{References}

Anderson, M. P., Characterization of geological heterogeneity,
in {\it Subsurface Flow and Transport: A Stochastic Approach},
Dagan, G. and S. P. Neuman (Eds.), Cambridge University Press,
New York,  23-43, 1997.

Barkai, E., and R. Silbey, Fractional Kramers equation,
{\it J. Phys. Chem.}, {\it 104}(16), 3866-3874, 2000.

Bear, J., {\it Dynamics of Fluids in Porous Media}, Elsevier,
New York, 1972.

Berkowitz, B., and H. Scher, On characterization of anomalous
dispersion in porous and fractured media, {\it Water Resour. Res.},
{\it 31}(6), 1461-1466, 1995.

Berkowitz, B., and H. Scher, Anomalous transport in random fracture
networks, {\it Phys. Rev. Lett.}, {\it 79}(20), 4038-4041, 1997.

Berkowitz, B., and H. Scher, Theory of anomalous chemical transport
in fracture networks, {\it Phys. Rev. E}, {\it 57}(5), 5858-5869, 1998.

Berkowitz, B., and H. Scher, The role of probabilistic approaches
to transport theory in heterogeneous media, {\it Transport Porous Media},
{\it 42}(1-2), 241-263, 2001.

Berkowitz, B., H. Scher, and S. E. Silliman, Anomalous transport in
laboratory-scale, heterogeneous porous media, {\it Water
Resour. Res.}, {\it 36}(1), 149-158, 2000.
[Minor correction: {\it 36}(5), 1371, 2000.]

Berkowitz, B., G. Kosakowski, G. Margolin, and H. Scher, Application
of continuous time random walk theory to tracer test measurements in
fractured and heterogeneous porous media, {\it Ground Water},
{\it 39}(4), 593-604, 2001.

Bouchaud, J.-P., and A. Georges,
Anomalous diffusion in disordered media - statistical mechanisms,
models and physical applications,
{\it Phys. Rep.}, {\it 195}(4-5), 127-293, 1990.

Chandrasekhar, S., Stochastic problems in physics and astronomy, {\it Rev.
Mod. Phys.}, {\it 15}(1), 1-89, 1943.

Compte, A., Stochastic foundations of fractional dynamics,
{\it Phys. Rev. E}, {\it 53}(4), 4191-4193, 1996.

Compte, A., Continuous time random walks on moving fluids,
{\it Phys. Rev. E}, {\it 55}(6), 6821-6831, 1997.

Compte, A., and M. O. C{\`a}ceres,
Fractional dynamics in random velocity fields,
{\it Phys. Rev. Lett.}, {\it 81}(15), 3140-3143, 1998.

Compte, A., R. Metzler, and J. Camacho,
Biased continuous time random walks between parallel plates,
{\it Phys. Rev. E}, {\it 56}(2), 1445-1454, 1997.

Cushman, J. H., X. Hu, and T. R. Ginn, Nonequilibrium statistical
mechanics of preasymptotic dispersion, {\it J. Stat. Phys.},
{\it 75}(5/6), 859-878, 1994.

Dagan, G., and S. P. Neuman (eds.), {\it Subsurface Flow and
Transport: A Stochastic Approach}, Cambridge University
Press, New York, 1997.

Dagan, G., Stochastic modeling of flow and transport: The broad
perspective, in  Dagan,  G. and S. P. Neuman (Eds.), {\it Subsurface
Flow and Transport: A Stochastic Approach}, 3-19, Cambridge
University Press, New York, 1997.

Deng, F.-W., J. H. Cushman, and J. W. Delleur, A fast Fourier transform
stochastic analysis of the contaminant transport problem,
{\it Water Resour. Res.}, {\it 29}(9), 3241-3247, 1993.

Eggleston, J., and S. Rojstaczer, Identification of large-scale
hydraulic conductivity trends and the influence of trends on
contaminant transport, {\it Water Resour. Res.}, {\it 34}(9), 2155-2168, 1998.

Feehley, C. E., C. Zheng, and F. J. Molz, A dual-domain mass
transfer approach for modeling solute transport in heterogeneous aquifers:
Application to the Macrodispersion Experiment (MADE) site,
{\it Water Resour. Res.}, {\it 36}(9), 2501-2515, 2000.

Garabedian, S. P., D. R. LeBlanc, L. W. Gelhar, and M. A. Celia,
Large-scale natural gradient tracer test in sand and gravel, Cape
Cod, Massachusetts 2. Analysis of spatial moments for a
nonreactive tracer,
{\it Water Resour. Res.}, {\it 27}(5), 911-924, 1991.

Gelhar, L. W., C. Welty, and K. R. Rehfeldt, A critical review
of data on field-scale dispersion in aquifers,
{\it Water Resour. Res.}, {\it 28}(7), 1955-1974, 1992.

Glimm, J., W. B. Lindquist, F. Pereira, and Q. Zhang,
A theory of macrodispersion of the scale-up problem,
{\it Transp. Porous Media}, {\it 13}, 97-122, 1993.

Gnedenko, B. V., and A. N. Kolmogorov, {\it Limit
Distributions for Sums of Random Variables\/}, Addison--Wesley,
Reading, 1954.

Haggerty, R., and S. M. Gorelick, Multiple-rate mass transfer for
modeling diffusion and surface reactions in media with pore-scale
heterogeneity, {\it Water Resour. Res.}, {\it 31}(10), 2383-2400, 1995.

Harvey, C., and S. M. Gorelick, Rate-limited mass transfer
or macrodispersion: Which dominates plume evolution at the
Macrodispersion Experiment (MADE) site?,
{\it Water Resour. Res.}, {\it 36}(3),
637-650, 2000.

Hatano, Y., and N. Hatano, Dispersive transport of ions in column
experiments: An explanation of long-tailed profiles,
{\it Water Resour. Res.}, {\it 34}(5), 1027-1033, 1998.

Hilfer, R., and L. Anton, Fractional master equations and
fractal time random walks, {\it Phys Rev. E}, {\it 51}(2), R848-R851, 1995.

Hilfer, R. (Ed.), {\it Applications of Fractional
Calculus in Physics}, 472 pp., World Scientific, Singapore, 2000.

Jespersen, S., R. Metzler, and H. C. Fogedby, L{\'e}vy flights in external
force fields: Langevin and fractional Fokker-Planck equations and their
solutions, {\it Phys. Rev. E}, {\it 59}(3), 2736-2745, 1999.

Kenkre, V. M., E. W. Montroll, and M. F. Shlesinger,
Generalized master equations for continuous-time random
walks, {\it  J. Stat. Phys.}, {\it 9}(1), 45-50, 1973.

Kinzelbach, W., {\it Groundwater Modelling}, Developments
in Water Science, 25, Elsevier, Amsterdam, 1986.

Klafter, J., and R. Silbey, On electronic energy transfer in disordered
systems, {\it  J. Chem. Phys.}, {\it 72}(2), 843-848, 1980a.

Klafter, J., and R. Silbey, Derivation of continuous-time random-walk
equations, {\it Phys. Rev. Lett.}, {\it 44}(2), 55-58, 1980b.

Klafter, J., A. Blumen, and M. F. Shlesinger,
Stochastic pathway to anomalous diffusion, {\it Phys. Rev. A},
{\it 35}(7), 3081-3085, 1987.

Koltermann, C. E., and S. M. Gorelick, Heterogeneity in sedimentary
deposits: A review of structure-imitating, process-imitating
and descriptive approaches, {\it Water Resour. Res.}, {\it 32}(9),
2617-2658, 1996.

Kosakowski, G., B. Berkowitz, and H. Scher, Analysis of field
observations of tracer transport in a fractured till,
{\it J. Cont. Hydrol.}, {\it 47}, 29-51, 2001.

Labolle, E. M., and G. E. Fogg,
Role of molecular diffusion in contaminant migration and
recovery in an alluvial aquifer system,
{\it Transp. Porous Media}, {\it 42}, 155-179, 2001.

Labolle, E. M., G. E. Fogg, and A. F. B. Tompson,
Random-walk simulation of transport in heterogeneous porous media:
Local mass-conservation problem and implementation methods,
{\it Water Resour. Res.}, {\it 32}(3), 583-593, 1996.

L{\'e}vy, P., {\it Calcul des probabilit{\'e}s\/},
Gauthier--Villars, Paris, 1925.

L{\'e}vy, P., {\it Th{\'e}orie de l'addition
des variables al{\'e}atoires\/}, Gauthier--Villars, Paris, 1954.

Li, S., and D. McLaughlin, Using the nonstationary spectral method
to analyze flow through heterogeneous trending media,
{\it Water Resour. Res.}, {\it 31}(3), 541-552, 1995.

Mantegna, R. N., and H. E. Stanley, Stochastic process with ultraslow
convergence to a Gaussian - The truncated L{\'e}vy flight,
{\it Phys. Rev. Lett.}, {\it 73}(22), 2946-2949, 1994.

Mantegna, R. N., and H. E. Stanley, Ultra-slow convergence to a
Gaussian: The truncated L{\'e}vy flight, in
{\it L{\'e}vy Flights and Related Topics in Physics},
edited by M. F. Shlesinger, G. M. Zaslavsky,
and U. Frisch, pp. 301-312, Springer-Verlag, New York, 1995.

Margolin, G., and B. Berkowitz, Application of continuous time random
walks to transport in porous media,
{\it J. Phys. Chem. B}, {\it 104}(16), 3942-3947, 2000.
[Minor correction: {\it 104}(36), 8762, 2000.]

Margolin, G., and B. Berkowitz, Spatial behavior of anomalous transport,
{\it Phys. Rev. E}, {\it 65}, in press, 2002.

Matheron, G., and G. de Marsily, Is transport in porous media
always diffusive? A counter example, {\it Water Resour. Res.},
{\it 16}(5), 901-917, 1980.

Metzler, R., Generalized Chapman-Kolmogorov equation: A unifying approach
to the description of anomalous transport in external fields,
{\it Phys. Rev. E}, {\it 62}(5), 6233-6245, 2000.

Metzler, R., and A. Compte,
Generalized diffusion-advection schemes and dispersive sedimentation:
A fractional approach, {\it J. Phys. Chem. B}, {\it 104}(16), 3858-3865, 2000.

Metzler, R., and J. Klafter, The random walk's guide to
anomalous diffusion: a fractional dynamics approach,
{\it Phys. Rep.}, {\it 339}(1), 1-77, 2000.

Metzler, R., J. Klafter, and I. M. Sokolov, Anomalous transport in
external fields: Continuous time random walks and fractional
diffusion equations extended, {\it Phys. Rev. E}, {\it 58}(2),
1621-1633, 1998.

Neuman, S. P., Eulerian-Lagrangian theory of transport in space-time
nonstationary velocity fields: Exact nonlocal formalism by
conditional moments and weak approximations, {\it Water Resour. Res.},
{\it 29}(3), 633-645, 1993.

Oldham, K. B., and J. Spanier, {\it The Fractional
Calculus\/}, Academic Press, New York, 1974.

Oppenheim, I., K. E. Shuler, and G. H. Weiss, {\it The Master Equation},
MIT Press, Cambridge, 1977.

Rubin, Y., Transport of inert solutes by groundwater: recent
developments and current issues, in
{\it Subsurface Flow and Transport: A Stochastic Approach}, Dagan, G.
and S. P. Neuman (eds.), 115-132, Cambridge University Press, Cambridge, 1997.

Samko, S. G., A. A. Kilbas, and O. I. Marichev, {\it Fractional Integrals and
Derivatives - Theory and Applications}, Gordon and Breach, New York, 1993.

Scheidegger, A. E., Statistical hydrodynamics in porous
media, {\it J. Appl. Phys.}, {\it 25}, 994-1001, 1954.

Scheidegger, A. E., Statistical approach to miscible
displacement in porous in porous media,
{\it Bull. Canad. Inst. Min. Met.}, 26-30, 1958.

Scher, H., and M. Lax, Stochastic transport in a disordered solid. I. Theory,
{\it Phys. Rev. B}, {\it 7}(10), 4491-4502, 1973.

Scher, H., and E. W. Montroll, Anomalous transit-time dispersion in
amorphous solids, {\it Phys. Rev. B}, {\it 12}(6), 2455-2477, 1975.

Scher, H., M. F. Shlesinger, and J. T. Bendler, Time-scale invariance
in transport and relaxation, {\it Physics Today}, January,
26-34, 1991.

Shlesinger, M. F., Asymptotic solutions of continuous-time
random walks,  {\it J. Stat. Phys.}, {\it 10}(5), 421-434, 1974.

Shlesinger, M. F., Random Processes, in {\it Encyclopedia of Applied
Physics}, Vol. 16, VCH Publishers, Inc., New York, 1996.

Shlesinger, M. F., G. M. Zaslavsky, and J. Klafter,
Strange kinetics, {\it Nature}, {\it 363}(6424), 31-37, 1993.

Silliman, S. E., and E. S. Simpson, Laboratory evidence of the scale
effect in dispersion of solutes in porous media, {\it Water Resour. Res.},
{\it 23}(8), 1667-1673, 1987.

Sposito, G., W. A. Jury, and V. Gupta, Fundamental problems in
the stochastic convection-dispersion model of solute transport
in aquifers and field soils,
{\it Water Resour. Res.}, {\it 22}(1), 77-88, 1986.

Zhang, Q., A multi-length-scale theory of the anomalous
mixing-length growth for tracer flow in heterogeneous
porous media,{\it J. Stat. Phys.}, {\it 66}(1/2),
485-501, 1992.

Zumofen, G., and J. Klafter,
Spectral random-walk of a single molecule,
{\it Chem. Phys. Lett.}, {\it 219}(3-4), 303-309, 1994.

Zumofen, G., J. Klafter, and A. Blumen, Trapping aspects in enhanced
diffusion, {\it J. Stat. Phys.}, {\it 65}(5/6), 991-1013, 1991.

\end{document}